# Epileptic Seizure Detection and Classification using Time-Frequency Features in EEG Signals

Abdullah Othman
Department of Electrical Engineering
King Fahd University of Petroleum and Minerals
Dhahran, Saudi Arabia

Mohamed A. Deriche
Department of Electrical Engineering
King Fahd University of Petroleum and Minerals
Dhahran, Saudi Arabia

*Abstract*— The use of EEG signal to diagnose several brain abnormalities is well-established in the literature. Particularly, epileptic seizure can be detected using EEG signals and several works were done in this field. The joint time-frequency domain features proved to be an improvement in classification results. This can be attributed to the non-stationarity inherent feature of the EEG signals. Therefore, shifting to the joint time-frequency domain benefits in utilizing the variations in time with respect to frequency and vice versa to obtain better outcomes. The Bayesian classifier is utilized to obtain the ranking of the best features. The information gain criterion was built to test the system and obtain the results.

## I. INTRODUCTION

Electroencephalogram (EEG) is a recording of the electrical activity in the brain detected from the scalp. The EEG signal is quite small in intensity, measured in microvolts (μV). EEG signals are used extensively in the diagnosis of brain-related abnormalities such as epileptic seizure [1]. Seizure can be described as a sudden surge of brain electrical activity which manifests itself in abnormal behavior, perception and thinking. Statistical studies, image and signal processing proved to be of high importance in classifying seizure and healthy EEG datasets [1, 2]. Analyzing EEG signals in time, frequency and joint time-frequency domains were introduced and studied in previous works such as [2, 3, 4, 5]. Typically, this step is the first step taken to analyze and classify EEG signals. Then, a few of these features that describe and characterize the seizure activity is selected. Finally, these features are classified using a certain classifier type to correctly detect the presence of epileptic behavior in the EEG dataset. The result of the classification is either a healthy or an abnormal seizure class. The features obtained in the time domain for example include statistical information such as the statistical mean and variation of the EEG segment. Time domain features also include amplitude-based features such as the median absolute deviation, and entropy-based features like Shannon entropy. Frequency-based features include: spectral-based features (such as the spectral flux, spectral centroid, spectral roll-off and spectral flatness). In the joint time-frequency domain, several features were considered that are like the time domain features e.g. mean, variance, inter-quartile range etc., while some new characteristics were considered e.g. the instantaneous frequency and frequency-dependent features. The importance of the joint time-frequency stems from the fact that EEG signals are non-stationary, that is, they are time-varying with frequency contents [2]. This gives more insight to the nature of EEG signals which means better selection and classification results in the joint time-frequency domain. Using Bayesian classifier, the set of features are then ranked to obtain the best set of features to reduce cost of computing and easiness of diagnosis method.

## II. TIME-FREQUENCY REPRESENTATIONS

EEG signals are inherently non-stationary. Therefore, transferring to time-frequency domain gives more insight into the nature of the EEG signal. Three quadratic time-frequency distributions are considered in-depth. To be specific, Smoothed Wigner-Ville Distribution (SWVD), Choi-Williams Distribution (CWD) and Spectrogram (SPEC) are used. The quadratic time-frequency distribution is given by:

$$\rho[n,k] = 2 \underset{n \to k}{DFT} \left\{ G[n,m] \underset{n}{*} (z[n+m]z*[n-m]) \right\} \quad (1)$$

Where $G$ is the time-lag kernel of the TFD and $*$ is convolution. $\rho[n,k]$ is an N x M matrix where M is the number of FFT points used. $z[n]$ is the analytic version of the real signal $x[n]$. $n = t.f_s$ and $k = \frac{2M}{f_s}f$ where $t$ and $f$ are the continuous time and frequency variables, and $f_s$ is the sampling frequency of the signal. Spectrogram is described by the following formula:

$$SP_x(t,f) = |ST_x(t,f)|^2 = ST_x(t,f)ST_x^*(t,f)$$
$$ST_x(t,f) = \int_{-\infty}^{\infty} x(\tau)\omega^*(t-\tau)e^{-j2\pi f\tau}d\tau \quad (2)$$

Choi-Williams distribution is described by the following formula:

$$C_x(t,f) = \int_{-\infty}^{\infty}\int_{-\infty}^{\infty} A_x(\eta,\tau)\Phi(\eta,\tau)\exp(j2\pi(\eta t - \tau f))d\eta d\tau$$

$$A_x(\eta,\tau) = \int_{-\infty}^{\infty} x(t+\tau/2)x^*(t-\tau/2)e^{-j2\pi t\eta}dt$$

and the kernel function is $\Phi(\eta,\tau) = \exp[-\alpha(\eta\tau)^2]$

Thirdly, the Smoothed Wigner-Ville is described by the following formula:

$$PW_x(t,f) = \int_{-\infty}^{\infty} \omega(\tau/2)\omega^*(-\tau/2)x(t+\tau/2)x^*(t-\tau/2)e^{-j2\pi\tau f}d\tau$$

### III. T-F ANALYSIS FOR EEG SEIZURE DETECTION

The T-F representation offers the starting and end frequency as well as the time range of signal components. Two samples of EEG signal for seizure and healthy segments are shown in figure 1. This shows how QTFD plot inform about the non-stationarity feature of the joint T-F representation as opposed to the traditional time or frequency representations. Figure 2 shows relevant image plots (Greyscale) for the QTFD representations.

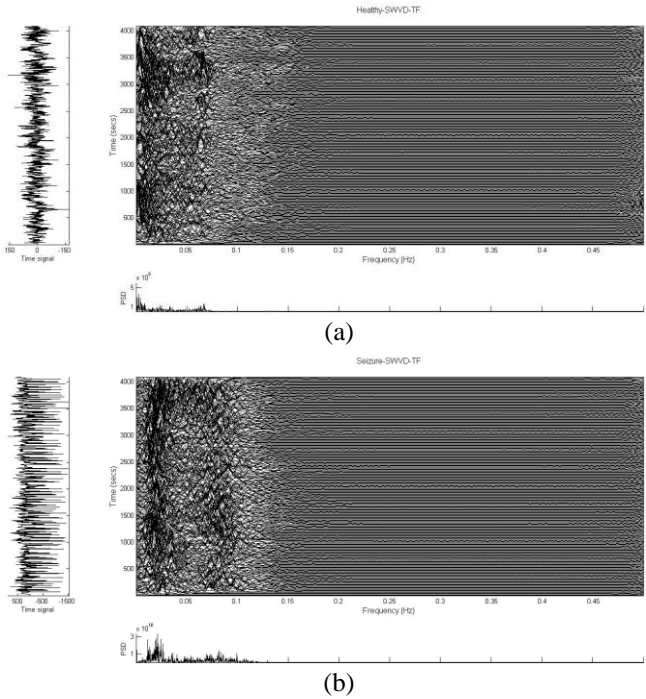

Fig. 1. (a) Healthy and (b) seizure activities QTFD generated using SWVD

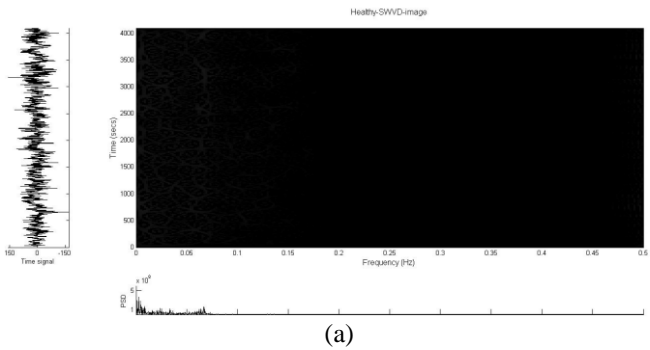

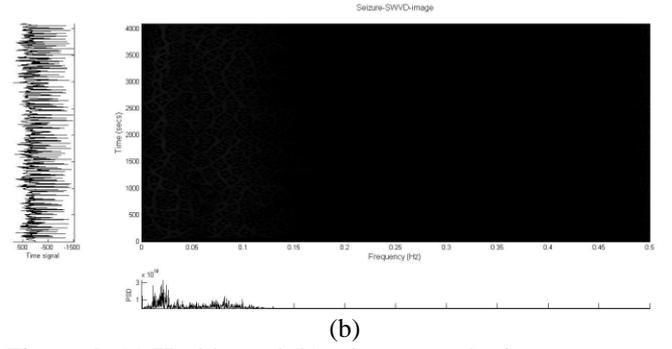

(b)

Figure. 2. (a) Healthy and (b) seizure samples image (greyscale)

### IV. EEG CLASSIFICATION USING BAYESIAN APPROACH

Based on [2], several time, frequency, time-frequency time and time-frequency frequency features are proposed for extracting the information needed for the classification from both healthy and seizure segments. Using Bayesian classifier, the percentage of accuracy in detecting different classes is put to test and comparison for the various systems.

### V. EEG FEATURE SELECTION AND EXTRACTION METHOD

Four types of features are chosen and described for use in extracting information of EEG signals. Tables 1 and 2 present the time-domain and frequency-domain features that are utilized in this experiment. The time-domain features include statistical-based and entropy-based features. The frequency-domain features include power spectrum-based and entropy-based features. Note that for the real discrete EEG signal $x$ then $z$ and $\rho$ are the analytical signal and the QTFD of the analytical signal, respectively. The analytical signal is obtained using the Hilbert transform.

TABLE 1. Time-Domain Features(F(t))

| Mean, Variance, Coefficient of Skewness, and Coefficient of Kurtosis: $$TiF_1 = \mu = \frac{1}{N}\sum_{n=1}^{N}|z[n]|$$ $$TiF_2 = \sigma^2 = \frac{1}{N}\sum_{n=1}^{N}(\mu - |z[n]|)^2$$ $$TiF_3 = \frac{1}{N\sigma^3}\sum_{n=1}^{N}(|z[n]| - \mu)^3$$ $$TiF_4 = \frac{1}{N\sigma^4}\sum_{n=1}^{N}(|z[n]| - \mu)^4$$ |
|---|
| Coefficient of variation $$TiF_5 = \frac{\sigma}{\mu} = \frac{\sqrt{TiF_2}}{TiF_1}$$ |
| Mean and median absolute deviations $$TiF_6 = Mean\{|\underline{z} - Mean(\underline{z})|\}$$ $$TiF_7 = Median\{|\underline{z} - Median(\underline{z})|\}$$ |

| Root Mean square (RMS) amplitude $$TiF_8 = \sqrt{\frac{1}{N}\sum_{n=1}^{N} z[n]^2}$$ |
|---|
| Inter-quartile range (IQR) $$TiF_9 = z\left[\frac{3(N+1)}{4}\right] - z\left[\frac{N+1}{4}\right]$$ |
| Entropy-based feature: Shannon entropy $$TiF_{10} = -\sum_{n=1}^{N} z[n]\log_2(z[n])$$ |

TABLE 2. TF-Translated Features (T(tf))

| $$TiTF_1 = \frac{1}{NM}\sum_{k=1}^{M}\sum_{n=1}^{N} \rho[n,k]$$ $$TiTF_2 = \frac{1}{MN}\sum_{k=1}^{M}\sum_{n=1}^{N} (\rho[n,k] - TiTF_1)^2$$ $$TiTF_3 = \frac{1}{(NM-1)TiTF_2^{3/2}}\sum_{k=1}^{M}\sum_{n=1}^{N} (\rho[n,k] - TiTF_1)^3$$ $$TiTF_4 = \frac{1}{(NM-1)TiTF_2^{4/2}}\sum_{k=1}^{M}\sum_{n=1}^{N} (\rho[n,k] - TiTF_1)^4$$ $$TiTF_5 = \frac{\sqrt{TiTF_2}}{TiTF_1} = \frac{\sigma_{(t,f)}}{\mu_{(t,f)}}$$ $$TiTF_6 = \text{Mean}\{|\underline{\rho} - \text{Mean}(\underline{\rho})|\}$$ $$= \frac{1}{NM}\sum_{k=1}^{M}\sum_{n=1}^{N} |\rho[n,k] - \mu_{(t,f)}|$$ $$TiTF_7 = \text{Median}\{|\underline{\rho} - \text{Median}(\underline{\rho})|\}$$ $$= \sqrt{\frac{\sum_{k=1}^{M}\sum_{n=1}^{N} \rho[n,k]}{NM}}$$ |
|---|
| $$TiTF_8 = \frac{1}{M}\sum_{k=1}^{M}\left(\rho\left[\frac{3(N+1)}{4}, k\right] - \rho\left[\frac{N+1}{4}, k\right]\right)$$ |
| $$TiTF_9 = -\sum_{n=1}^{N}\sum_{k=1}^{M} \rho[n,k]\log_2(\rho[n,k])$$ |

TABLE 3. TF-Translated Features (F(f))

| Power spectrum-based feature: maximum power of the frequency bands |
|---|
| $$FrF_1 = \sum_{k=1}^{\delta} |Z[k]|^2$$ $$FrF_2 = \sum_{k=\delta+1}^{M} |Z[k]|^2$$ |
| Spectral-based features:<br>• Spectral flux: difference between normalized spectra magnitudes $$FrF_3 = \sum_{k=1}^{M} \left(Z^{(l)}[k] - Z^{(l-1)}[k]\right)^2$$ $Z^l$ and $Z^{l-1}$ are normalized magnitude of the Fourier transform at L and L $-$ 1 frames<br>• Spectral centroid: average signal frequency weighted by magnitude of spectral centroid $$FrF_4 = \frac{\sum_{k=1}^{M} k|Z[k]|}{\sum_{k=1}^{M} |Z[k]|}$$<br>• Spectral Roll-Off (i.e. spectral concentration below threshold) $$FrF_5 = \lambda \sum_{k=1}^{M} |Z[k]|$$ In this study, $\lambda$ is chosen to be 0.85 (frequency under which 85% of the signal power resides)<br>• Spectral flatness: indicates whether the distribution is smooth or spiky $$FrF_6 = \left(\prod_{k=1}^{M} Z[k]\right)^{\frac{1}{M}} \left(\sum_{k=1}^{M} |Z[k]|\right)^{-1}$$ |
| Spectral entropy-based feature: measure the regularity of the power spectrum of the EEG signal [8] $$FrF_7 = \frac{1}{\log(M)}\sum_{k=1}^{M} P(Z[k])\log P(Z[k])$$ |

TABLE 4. TF-Translated Features (F(tf))

| $$FrTF_1 = \sum_{n=1}^{N}\sum_{k=1}^{M_\delta} \rho[n,k]$$ $$FrTF_2 = \sum_{n=1}^{N}\sum_{k=M_\delta+1}^{M} \rho[n,k]$$ } Sub-band energies |
|---|
| where $M_\delta = \lfloor \frac{M}{f_s} \rfloor$ |
| $FrTF_3 = \sum_{n=1}^{L} \sum_{k=1}^{P} (\rho[n,k] - \rho[n+L,k])^2$ <br> L is a predetermined lag and P is the total of sub-bands <br> $FrTF_4 = \frac{\sum_{k=1}^{M} k\rho[n,k]}{\sum_{k=1}^{M} \rho[n,k]}$ ($\simeq$ instantaneous frequency) <br> $FrTF_5 = \lambda \sum_{n=1}^{N} \sum_{k=1}^{M} \rho[n,k]$ <br> $FrTF_6 = \frac{(\prod_{n=1}^{N} \prod_{k=1}^{M} \rho[n,k])^{\frac{1}{NM}}}{\sum_{k=1}^{M} \sum_{n=1}^{N} \rho[n,k]}$ <br> ($\simeq$ ENERGY LOCALIZATION) |
| $FrTF_7 = \frac{1}{1-\alpha} \log_2 \left( \sum_{n=1}^{N} \sum_{k=1}^{M} \rho^\alpha[n,k] \right)$ |

## VI. DATABASE AND RESULTS

The database chosen for conducting the research was first analyzed in [9]. It consists of five sets each containing 100 single-channel EEG segments of duration 23.6 seconds. Sets A and B consisted of segments obtained from surface EEG recordings that were carried out on five healthy volunteers using standardized electrode placement scheme. For (A) the volunteers were relaxed with eyes open and in (B) with eyes closed. Sets (C), (D) and (E) originated from an EEG archive of pre-surgical diagnosis. Five patients' data were selected, all of whom had complete seizure control after resection of one of the hippocampal formations. Segments in (D) were taken from the epileptogenic zone, while those in (C) were recorded from the hippocampal formations of the opposite hemisphere of the brain. Both (C) and (D) were measured during seizure-free intervals. Group (E) contains data recorded during seizure activity. All EEG signals were recorded using 128 channel amplifier system, using an average common reference. Digital to analog conversion of 12 bits was implemented, and the data were sampled onto the disk of a data acquisition computer at a rate of 173.6 Hz. Low-pass filter of 40 Hz was applied. For the T-F features, the FFT window length is set to 512 samples. The lag window is set to a length of 127. To evaluate the performance of the features, a Bayesian classifier of type Naïve Bayesian is prepared. Two classes constitute the new database, where one class consists of 100 segments from set (A). The second class consists of 100 segments from set (E). To train the classifier, 30% of the data i.e. 60 segments for the two types were fed to the classifier for training. 70% were used to test the classifier performance. Table 5 shows the total classification accuracy results by combining the time domain and the frequency domain features, and then by combining the respective features from the joint time-frequency features for different QTFDs. The result is an improvement by almost 3.5% if all features are taken. The classification was done using the Naïve Bayes classifier. A set of four best features was taken using the information gain criterion ranker and it shows an improvement by about 9% for the T-F translated features. This step gives an advantage of reducing the computation cost by evaluating four features instead of 16. There is also a noticeable improvement in the classification results by using the T-F features. Table 6 shows the ranking of the T-F features to understand the relevance of these features. This ranking is achieved using the information gain criterion.

TABLE 5. TOTAL CLASSIFICATION ACCURACY RESULTS

| Features | Total classification accuracy | | |
|---|---|---|---|
| {F(t), F(f)} | 96.43% | | |
| *{TiF1, TiF9, TiF2, FrF2}* | 90.18% | | |

| Features | Total classification accuracy | | |
|---|---|---|---|
| | SWVD | CWD | SPEC |
| F(tf) = {F(t), F(f)} | 100% | 100% | 100% |
| *{TiTF1, TiTF8, TiTF2, FrTF11}* | 99.286% | 99.67% | 99.13% |

TABLE 6. RANKING OF THE T-F FEATURES BASED ON THE INFORMATION GAIN CRITERION

| Rank | SWVD | CWD | SPEC |
|---|---|---|---|
| 1 | $TiTF_1$ | $TiTF_1$ | $TiTF_1$ |
| 2 | $TiTF_8$ | $TiTF_8$ | $TiTF_8$ |
| 3 | $TiTF_2$ | $TiTF_2$ | $TiTF_2$ |
| 4 | $FrTF_2$ | $FrTF_2$ | $FrTF_2$ |
| 5 | $TiTF_6$ | $TiTF_6$ | $TiTF_6$ |
| 6 | $FrTF_5$ | $FrTF_5$ | $FrTF_5$ |
| 7 | $FrTF_4$ | $FrTF_4$ | $FrTF_4$ |
| 8 | $TiTF_9$ | $TiTF_7$ | $TiTF_9$ |
| 9 | $TiTF_5$ | $TiTF_5$ | $TiTF_5$ |
| 10 | $TiTF_7$ | $TiTF_9$ | $TiTF_7$ |
| 11 | $FrTF_1$ | $FrTF_1$ | $FrTF_1$ |
| 12 | $FrTF_3$ | $FrTF_3$ | $FrTF_3$ |
| 13 | $FrTF_7$ | $FrTF_7$ | $FrTF_7$ |
| 14 | $TiTF_4$ | $TiTF_3$ | $TiTF_3$ |
| 15 | $FrTF_6$ | $FrTF_6$ | $FrTF_6$ |
| 16 | $TiTF_3$ | $TiTF_4$ | $TiTF_4$ |

Figures 3 and 4 show the histogram and the fitting probability curves for two selected features. Figure 3 shows the result for *TiTF1*, the mean. Figure 4 shows the result for *TiTF3*. It can be inferred that since in figure 3 there is a small interference between the seizure case (red) and the normal case (green) normal distribution curves, the classification result is good comparatively to figure 4 where the interference is higher.

## CONCLUSION

Using joint time-frequency features improved the efficiency of EEG detection and diagnosis by about 9% for the best features, and by about 3% for all the features combined.

Hence, going to the joint time-frequency domain reduces computational cost and improves the accuracy of diagnosis for epileptic suspected cases. Furthermore, it was found that using Choi-Williams distribution was optimal among the QTFDs as it achieved better result. The normal distribution curves were used as a visual tool to differentiate between the features in correct diagnosis.

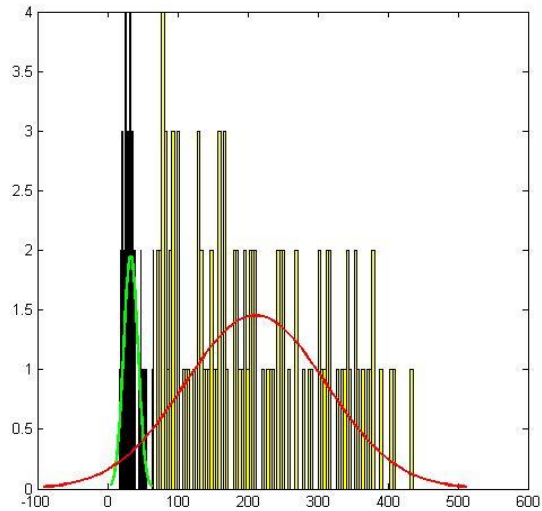

Figure 3: Histogram and probability curves for feature *TiTF1*, the mean

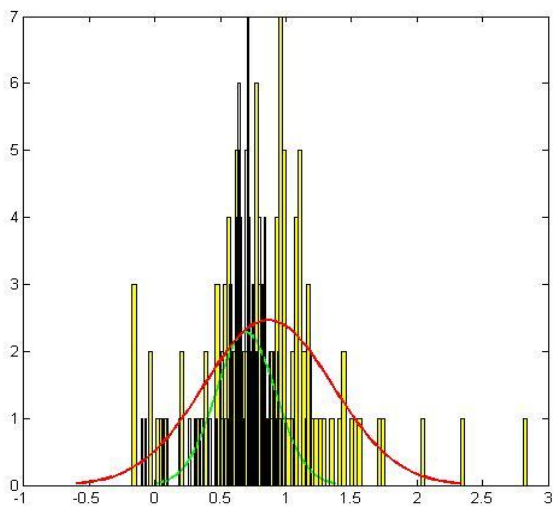

Figure 4. Histogram and Probability Curve for the skewness feature *TiTF3*